\documentclass[aps,twocolumn,floats,prd,psfig]{revtex4}
\usepackage{graphicx, epsfig}

\newcommand{\ApJL}{Astrophys. J. Lett.}
\newcommand{\ApJ}{Astrophys. J.}
\newcommand{\PRL}{Phys. Rev. Lett.}
\newcommand{\PRD}{Phys. Rev. D}

\newcommand{\MNRAS}{Mon. Not. Roy. Astron. Soc.}

\newcommand{\sun}{\odot}
\newcommand{\so}{M_\odot}
\newcommand{\beq}{\begin{equation}}
\newcommand{\eeq}{\end{equation}}
\newcommand{\beqa}{\begin{eqnarray}}
\newcommand{\eeqa}{\end{eqnarray}}

\newcommand{\lsim}{\lesssim}
\newcommand{\gsim}{\gtrsim}

\newcommand{\lmk}{\left(}
\newcommand{\rmk}{\right)}

\newcommand{\mb}{M_\bullet}

\begin{document}

\title{Search for Small-Mass Black Hole Dark Matter with 
Space-Based Gravitational Wave Detectors}
\author{Naoki Seto and Asantha Cooray}  
\affiliation{Theoretical Astrophysics, Division of Physics,
Mathematics and Astronomy, California Institute of Technology, MC
130-33, Pasadena, CA 91125; (seto,asante)@caltech.edu} 


\begin{abstract}
The high sensitivity of upcoming space-based
 gravitational wave detectors suggests the possibility that if halo dark 
matter were composed of primordial black holes (PBHs) with mass
between $10^{16}$ g and 10$^{20}$ g, 
the gravitational interaction with detector test masses
will lead to a detectable pulse-like signal during the fly-by.
For an improved version of the Laser Interferometer Space Antenna with a
reduced acceleration noise at the low-end of its frequency spectrum, we find 
an event rate, with signal-to-noise ratios greater than 5, of $\sim$ a 
few per decade involving black holes of mass $\sim$ 10$^{17}$ g.  
The detection rate improves significantly for second generation space based interferometers
that are currently envisioned, though these events must be
distinguished from those involving perturbations due to near-Earth asteroids.
While the presence of primordial black holes below a mass of $\sim$ 10$^{16}$
g is now constrained based on the radiation released during their evaporation,
the gravitational wave detectors will extend the study of PBHs to a several orders of magnitude
higher masses.
\end{abstract} 

\maketitle



Based on observed rotational velocity measurements of the Milky Way disk, 
the presence of a substantial dark matter component in the Milky Way 
halo is now well established 
\cite{Faber}. With no unique guidance as to the nature of dark matter,
a large number of candidates based on both astronomical arguments,
 such as baryonic dark matter involving substellar mass remnants \cite{Tri87}, and 
particle physics expectations \cite{Jun95} are now routinely considered to explain
the missing mass. An interesting possibility is that the halo dark matter is 
composed of primordial black holes (PBHs)  \cite{Hawking}. While
 PBHs are expected over a rather wide range of masses \cite{Nov79},
the population below $\sim 5 \times 10^{14}$ g
is expected to have evaporated by the Hawking radiation over the
age of the universe \cite{Haw74}, while holes with a mass 
slightly above this limit are still emitting high energy particles today \cite{Mac91}.
With observations related to the galactic Gamma-ray background, 
the evaporating black hole contribution to the
dark matter density in the Universe, in terms of the critical density, 
is at the level roughly below 10$^{-8}$ \cite{Mac91}.

From the high mass end, constraints on the possibility that dark matter is 
composed of PBHs come from dynamical arguments such as the potential disruption of galactic 
clusters and similar bound structures \cite{CarSak99}. These constraints generally 
limit the primordial black hole mass to be roughly below few solar mass. Similar constraints come from galactic microlensing 
observations, which limits  primordial black hole masses to be below roughly few tenths solar masses \cite{Alcock}.
In combination, 
at least over a 10$^{16}$ decade in primordial black hole mass, between evaporating holes with mass below 10$^{17}$ g 
and $\sim$ 0.1 M$_{\sun}$, remains yet to be studied and techniques to
understand their presence are limited (see also \cite{gwr}).
Though the galactic gamma-ray background and its weak anisotropy suggests 
a evaporating PBH density of roughly 10$^{10}$ pc$^{-3}$, PBHs can
explain the total halo dark matter if the PBH mass spectrum enhances the abundance above 10$^{17}$ g \cite{Cli98}. 
A promising possibility of probing small-mass PBHs involves  
femtolensing of gamma-ray bursts (GRBs) both in terms of an interference pattern in
the frequency spectrum \cite{Gou92} or relative flux differences in Astronomical Unit (AU)-scale separated light curves 
\cite{NemGou95}. The existing data weakly constrain the
mass density fraction, relative to the critical density, of compact objects with mass between 10$^{17}$ g and
10$^{20}$ g to be below 0.2, if the average redshift of GRBs is 
unity \cite{Marani}. This technique only constrain the extragalactic density, though,
it would be interesting to see if the prediction \cite{Cli98} related to the galactic dark matter can directly be tested. 

The PBHs, especially at the small-mass end,  are not expected to interact or be captured by other massive bodies such as the
Sun; they can be considered as another weakly interacting massive
particle (WIMPs). In contrast to active PBHs with masses
$\lsim 10^{15}$g,  the physical nature of  PBHs with  $\gsim 10^{17}$g is
characterized only by their masses  and extremely small size. 
In reality, they can be detected only through the
gravitational interaction. But, as their masses are much larger than standard WIMPs
predicted by particle physics, the expected flux of PBHs 
would be very small and we need detectors with a large effective area for their search.  

In this {\it Letter}, we suggest that PBHs around a mass of 10$^{17}$ g could in fact be detected
directly with planned space-based gravitational wave detectors that have
large cross sections to moving massive bodies. These missions include
the Laser Interferometer Space Antenna (LISA) \cite{Bender} 
and several vision missions of future that are now routinely discussed in the long term planning by various space
agencies: Gravitational Echos Across Time (GREAT; \cite{Cornish}), the Big Bang Observer (BBO) of NASA's Structure and
Evolution of the Universe Roadmap, and
a similar mission concept named Decihertz Interferometer Gravitational
Wave Observatory (DECIGO; \cite{Seto01}) (see also \cite{vec,Bender:vw}). 
The direct detection simply involves the gravitational interaction between the fly-by PBHs and detector test masses
such that with high sensitivity even a small gravitational perturbation can eventually be seen above the detector noise.
As an example, the LISA mission is expected to  probe the frequency range between 10$^{-5}$ Hz and 10$^{-1}$ Hz
 and detect perturbations above $h \sim \delta
L/L \sim 10^{-21}$ at $\sim 1$mHz
with detectors separated by arm lengths $L \sim 5 \times 10^6$ km. 

The interaction of a PBH and a gravitational-wave observatory test mass can be
considered either as a direct interaction, when the fly-by separation is smaller than the arm length,
or as a tidal interaction, in the case where arm length is smaller than the fly-by distance. We first 
consider the  former situation and take the scenario with a PBH mass $\mb$ passing by  one of
the test masses of the interferometer with a relative velocity $V$ with the closest approach distance $R$. 
The time dependent acceleration $a(t)$ of the spacecraft test mass towards  the direction of the closest approach has a single
pulse-like structure  given by
\beq
a(t)=\frac{G\mb R}{[R^2+(Vt)^2]^{3/2}},\label{at}
\eeq 
where we have set the origin of the time coordinate, $t=0$, to be at
the closest approach.  Note that the fly-by PBH can also perturb other test masses of the interferometer configuration since
at least 3 or more spacecrafts are generally utilized in future gravitational wave observatories. When making an
order-of-magnitude estimate we neglect additional information related to such perturbations and ignore
aspects that involve the relative configuration of interferometer-body system related to the perturbing PBH. 

The Fourier component $a(f)$ of eq.~(\ref{at}) is given in terms of the
1st order modified Bessel function $K_1$ as 
\beq
a(f)=\int^\infty_{-\infty} dt e^{2\pi if t}a(t)= 2\frac{G\mb}{RV} K_1[2\pi
fR/V]. 
\eeq 
For the direct detection, the signal-to-noise ratio (SNR) of the pulse $a(t)$ is formally
written as $SNR^2=2\int^\infty_{0} a(f)^2/a_n(f)^2df$ with the noise
spectrum $a_n(f)$ of the detector.  In the case of space-based
interferometers, the noise  has two main contributions involving the proof-mass
noise and the optical path noise with the former dominating the low frequency
regime. As we will soon discuss, the typical signal $a(f)$ due to the fly-by PBH of  mass around 
$\mb \gsim 10^{17}$ g has support in the SNR integral when $f\lsim 10^{-4}$ Hz. This is in the
the low frequency  regime of all the proposed future interferometers, including the LISA mission. The proof-mass
noise of LISA is estimated to be constant down to $\sim 10^{-4}$ Hz; It 
might become a factor of two larger at $3\times 10^{-5}$ Hz and more at lower frequencies  \cite{Bender}, 
however, the low frequency behavior of the noise curve, around $10^{-5}$ Hz, is yet to be determined
precisely.  In order to estimate the SNR, we simply extrapolate the constant proof-mass noise down to $f=0$,
and set $a_n(f)=2 a=const$,  where $a$ is the proof-mass noise for a single inertial
sensor in various interferometers. While frequency dependence of noise below $10^{-4}$ Hz is ignored, we 
do pay attention to the frequency dependence of the fly-by pulse event so as to 
understand requirements on the low frequency behavior of the proof-mass noise.

After some straight forward algebra, we obtain the SNR as
\beq
SNR^2=\frac{3\pi}{32} \frac{(G\mb )^2}{VR^3 a^2}~~~~(R<L).\label{snr0}
\eeq 
Note that this relation corresponds  to the second formula of eq.~(4.35) in Ref.~\cite{Bender}, but our result
differs by $1/13 \sim 8\%$  due to the exact use of a limit associated with the Bessel function,
instead of an approximation used there, and after taking in to account a difference in the definition related to 
the proof-mass noise.

Since the above expression~(\ref{snr0}) is for the case where the closest  approach
distance $R$ to a single spacecraft is smaller than the arm-length $L$ of the
interferometer, namely $R<L$, we now consider the opposite case with  $R>L$. Now, the tidal  deformation of the
interferometer is a measurable effect, but this involves 
an additional suppression factor $\sim n (L/R)$ in eq.~(\ref{at}), where we have ignored the numerical prefactor $n$
as we do not consider detailed geometry of these events, though we note that this factor 
could be greater than unity (up to 2) in certain favorable configurations of the interferometer with
respect the trajectory of the PBH fly-by. In this tidal limit, we find the SNR to be
\beq
SNR^2=\frac{3\pi}{32} \frac{(G\mb L)^2}{VR^5 a^2}~~~~(R>L). \label{sn}
\eeq
This is the relevant expression for LISA and BBO/DECIGO in the interesting
mass range $\mb\gsim 10^{17}$g, though, at the low mass end (10$^{16}$ g), the former expression applies
 for LISA.

Using eq.~(\ref{sn}) we can now estimate the maximum length of the closest
approach $R_{\rm max}$ for a given SNR  threshold 
\beqa
R_{\rm max}&=&5.3\times 10^{11}  \lmk \frac{\mb}{10^{17}{\rm g}}  \rmk^{2/5}  \lmk\frac{V}{\rm 350 km/s}  \rmk^{-1/5}\nonumber\\
& &\times  \lmk
\frac{SNR}5\rmk^{-2/5} \lmk \frac{a}{a_0}  \rmk^{-2/5} \lmk\frac{L}{L_0}
\rmk^{2/5} {\rm cm}
\eeqa
where
$a_0=3 \times 10^{-15}{\rm m/s^2\sqrt{Hz}}$ and  $L_0=5 \times
10^{11}$cm are the reference fiducial parameters for LISA in its current design \cite{Bender}
(at least  the proof-mass noise around $f\sim 10^{-4}$Hz).  We have taken the typical velocity
dispersion $V=220\sqrt{5/2}=350$ km/s of the halo dark matter particles
relative to  the solar system using the Galactic rotation velocity of
$220$ km/s  and the Galactic  radius to the Solar system of $r_g\sim 8$ kpc following Ref.~\cite{CarSak99}.
The gravitational perturbation involves a pulse like signal in the data streams with a
characteristic frequency given by
\beqa
V/R_{\rm max}&\sim&6.4\times 10^{-5}  \lmk \frac{\mb}{10^{17}{\rm g}}  \rmk^{-2/5}  \lmk\frac{V}{\rm 350 km/s}  \rmk^{6/5}\nonumber\\
& &\times  \lmk
\frac{SNR}5\rmk^{2/5} \lmk \frac{a}{a_0}  \rmk^{2/5} \lmk\frac{L}{L_0}
\rmk^{-2/5} {\rm Hz}
\eeqa
As mentioned already,  for most planned space interferometer such as LISA,
such a low frequency is in the regime where the detection is limited by the
proof-mass noise.

When we assume that the halo dark matter with density $\rho_{DM}\sim
0.011\so{\rm /pc^3}$ \cite{Olling} around
the Sun is made with PBHs of mass $\mb$,  their flux, $F=\rho_{DM} V  \mb^{-1}$, is 
\beqa 
F&=&9.0\times 10^{-27} \lmk\frac{\rho_{DM}}{0.011{\rm M_{\odot} pc^{-3}}}
\rmk \lmk\frac{V}{\rm 350 km/s}  \rmk \nonumber\\
& & \times \lmk \frac{\mb}{10^{17}{\rm g}}  \rmk^{-1} {\rm cm^{-2} yr^{-1}},
\eeqa
and the fly-by event rate, $\dot{\eta} = \pi F R_{\rm max}^2$, above a certain SNR is
\beqa
&&\dot{\eta}=0.01  
\lmk \frac{\mb}{10^{17}{\rm g}}  \rmk^{-1/5} \lmk \frac{SNR}5\rmk^{-4/5} 
\lmk\frac{V}{\rm 350 km/s}  \rmk^{3/5} \nonumber\\
& &\times    \lmk\frac{\rho_{DM}}{0.011{\rm M_{\odot} pc^{-3}}} \rmk
 \lmk \frac{a}{a_0}  \rmk^{-4/5} \lmk\frac{L}{L_0}
\rmk^{4/5}  \lmk\frac{N}1  \rmk {\rm yr^{-1}},\label{rate} \nonumber \\
\eeqa
where $N$ represents the effective number of interferometers with
sufficient relative distances.  While LISA involves a single set of interferometers,
some planned missions, {\it e.g.} BBO, plans to use multiple
interferometer arrays with large separation to improve the localization of binaries sources such that
$N\ge 2$. If the signal is to  be extracted using  a correlation analysis, especially for low signal-to-noise events,
of two arrays, there is no further increase in the event rate and one should continue to use $N=1$. 

For the reference parameters of LISA, assuming a useful SNR threshold of 5,  the detection rate is $\sim
0.1$ per ten years as shown in eq.~(\ref{rate}).  In the tidal limit, the event rate weakly depends on mass $\mb$ by
$\dot{\eta} \propto \mb^{-1/5}$. The combination $(aL^{-1})$ that appears in
all of the above expressions determines the low frequency sensitivity of the interferometers to the  gravitational wave amplitude $h$. 
Therefore, once the threshold distances $R_{\rm max}$ are confirmed to  be larger than the arm-length
$L$,  we can easily compare the event rate for various
interferometers by studying their noise curves  at the low frequency regime. 
In the case of LISA,  when the mass $\mb$ falls below $10^{17}$ g, the maximum fly-by distance approaches that of the
arm length ($\sim 5 \times 10^{11}$ cm). In this case, we repeat the event rate using the SNR involving the direct perturbation
(eq~\ref{snr0}); in this limit, for reference, the event rate depends on the mass $\mb$ by $\dot{\eta}\propto
\mb^{1/3}$.

\begin{figure}[t]
\centerline{\psfig{file=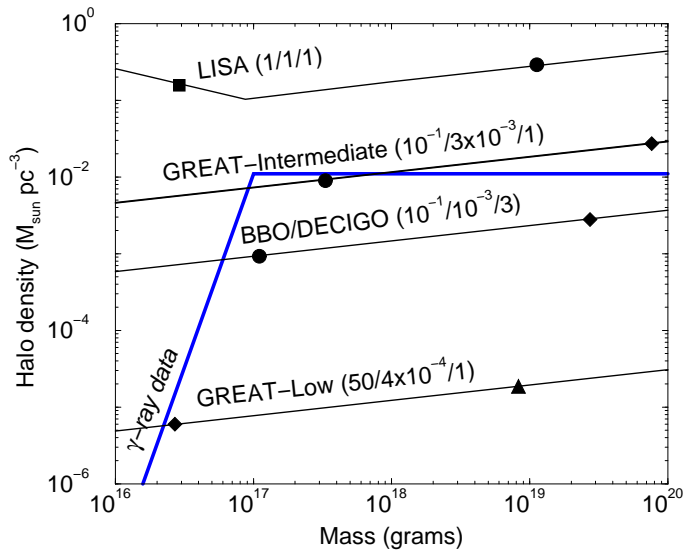,width=2.9in,angle=-90}}
\caption{The detection thresholds for PBH fly-by with various upcoming
 space-based   gravitational wave observatories (in solid lines
from top to bottom: LISA, GREAT intermediate-frequency-mission, BBO/DECIGO, and
GREAT very-low-frequency-mission). The thick solid
line shows the expected density of halo dark matter in the form of black 
holes; at $\mb \simeq 10^{15}$ g, the density is constrained by $\gamma$-ray background observations such that 
the particle density is below $\sim 10^{10}$ pc$^{-3}$,
while no similar constraints exist above $\mb\gsim 10^{17}$ g and we take
 the whole halo to be formed of black holes with the mass given in
the horizontal axis (with a density 0.011 M$_{\sun}$ pc$^{-3}$). 
The detection limits assume an event rate of a
 single  detection per decade; if no events are detected, these
curves would roughly correspond to the constraint one can put on the
 black  hole contribution to the halo dark matter density. In
each of the instruments considered, we label the following three 
parameters: (arm-length/proof-mass noise/total number of
detector arrays) relative to those reference values of LISA (arm length:
 $5\times 10^6$km, proof-mass noise: $3\times 10^{-15}{\rm m/s^2/\sqrt{Hz}}$). The symbols on each of 
the curves represent the average frequency of the gravitational perturbation produced on the
 interferometer  by a  black hole of corresponding mass with triangle,
diamond, circle and square representing frequencies of 
10$^{-7}$,10$^{-6}$,10$^{-5}$ and 10$^{-4}$ Hz, respectively. 
In the case of LISA, the threshold distance $R$ equals to the arm length $L$ at
 $\mb =0.87 \times 10^{17}$ g where the slope of the curve changes.}
\label{fig:limits}
\end{figure}

In figure 1, we show the potential detectability of fly-by events involving PBHs with mass $\mb$ using various planned space-based
interferometers. Instead of the event rate, our results are presented in terms of the local halo density of PBHs required for a
detection, with SNR greater than 5, of one event per decade.  For reference,  to put these
detections in the context of the galactic dark matter density, we also show current observational constraints for the
PBH density from the $\gamma$-ray background ($\mb\lsim 10^{17}$ g) \cite{Cli98} and  the expected density of  dark matter
based on a Galactic model ($\mb \gsim 10^{17}$g) where the 
density is now determined to be $(0.011 \pm 0.005)$ M$_{\sun}$ pc$^{-3}$ \cite{Olling}. In the case of LISA, the  
transition ($R=L$) between eqs.~(3) and (4) occurs at  $\mb =0.87\times
10^{17}$g, but for all other interferometers, we find $R>L$ in the mass range considered  in figure 1.

In figure~1, next to the name of each of these interferometers, we also denote three 
basic numbers that characterize an interferometer
(arm-length/proof-mass noise/number of detector arrays) normalized by
the reference values for LISA.  The GREAT-intermediate mission \cite{Cornish} has a
marginal possibility for a PBH detection at $\mb\sim 10^{17}$g while The GREAT-very low frequency mission
with parameters $(50/4 \times 10^{-4}/1)$ would detect them even if less than a few percent of the halo dark
matter is made with PBHs with mass between $10^{17}$ g and $10^{20}$ g.  
Incidently, the low-frequency end of the LISA noise has not been finalized and there are
strong scientific reasons to improve it beyond what was originally suggested \cite{Bender03}.
If we take the most improved low-frequency noise from Ref.~\cite{Bender03}
with scaled parameters (5/0.1/1), we find a detection level
similar to that shown for the GREAT-intermediate mission suggesting that even with an improved version of LISA,
a detection with an event rate of few per decade, at mass of $\mb \sim 10^{17}$ g, may be within the reach of
first-generation interferometers. We note here that if the event rate is too high with $\gsim 100 \; {\rm yr^{-1}}$, 
as would be the case for GREAT-low frequency mission \cite{Cornish}, 
the  signals will be overlapped in frequency space due to the low frequency nature.
This will complicate the extraction and would involve a detailed analysis similar to the ones
proposed to extract individual gravitational waves in their confusion limit.
In this figure, we also note the characteristic frequency of the fly-by event. This is done to emphasize the
very low frequency aspect of the detection, especially with respect to the noise requirements
associated with proof-mass, or acceleration noise.

In the search for fly-by PBHs, it is important to distinguish their
signals from other signals in the interferometer.  The primary targets of a space-based interferometer are low frequency
gravitational waves. In general, their measurement is limited to some extent by the astrophysical
confusion noise, depending on the frequency. For example, the LISA
detector noise is expected to be dominated by the Galactic
white dwarf binary confusion noise in the frequency band between $0.1\; {\rm mHz} \lsim f \lsim 3$ mHz \cite{Bender:vw}. At
a lower frequency than this range, gravitational waves from merging massive black hole binaries 
might form a confusion noise \cite{Bender:vw}. When we measure the local acceleration related to PBH fly-by events, however, we can
largely reduce the relative  contribution of the low frequency gravitational waves 
(with wave-length $\lambda_{GW} > L$) in the data streams by taking a certain
combination of the data stream now knowns as the Sagnac \cite{tinto}. This data combination nearly
cancels gravitational waves using the fact that they propagate with speed of
light, though it would still be affected by fly-by perturbations. 
Therefore, the separation of PBH signals with other gravitational
wave signals is not as severe as the  confusion noise problem for gravitational wave observations alone.  

There is one source of confusion, however. This involves similar pulses produced during 
the passage of near-Earth asteroids close to detectors (see, section 4.2.5 of \cite{Bender}). Given the expected
flux of minor bodies in the solar system, as determined by various observational data, the asteroid perturbations
are dominated by those at the high mass end between $10^{13}$ g to $10^{15}$ g,
or sizes around a km or slightly less,   
with an event rate of the order 0.05 yr$^{-1}$ (the rate for smaller mass events are substantially less and can be
ignored). Unfortunately, though the relative velocity of asteroid events are smaller ($\sim$ 30 km sec$^{-1}$) than
those involving PBHs,  these events could have similar frequencies (as PBHs) due to differences in the 
maximum distance to which they can be detected.  On the positive side, 
orbital parameters of roughly 10\% of such near-Earth asteroids are already known while this
fraction is soon expected to grow substantially with dedicated near-Earth asteroid search programs \cite{Stuart}
such that during the operation of these gravitational wave observatories, expected perturbations from asteroids can be
a priori determined. Moreover, we expect optical followup searches of all candidate events to further refine the sample
and to confirm the compactness of perturbers. Detection of a fly-by pulse with no
optical counterpart would be a minimum criteria for the selection of a
PBH event.  For  this perpose, we  need  a systematic optical  survey
around the interferometric area above some magnitude threshold. 
With increasing sensitivity, such as the case with GREAT-low frequency mission, event rate of
asteroids increases such that they might become the primary confusion noise; Removal of this confusion noise requires
independent estimates of the asteroid mass at the few percent level in addition to the orbit.

While with gravitational  wave detectors the presence of PBHs can be
established, it is not easy to determine masses of individual events from a single pulse signal
that is characterized mainly by two numbers: the amplitude, $M/R^3$, and
the time scale, $R/V$, made from  three variales $M$, $R$ and $V$. 
This is due to the fact that there is an unknown associated with the
fly-by distance or the PBH trajectory. If multiple detectors are perturbed, one can establish the trajectory
and then used that information to determine mass. 

While there is some possibility for PBH studies with LISA, the
possibilities are greater with second generation space-based missions for which concept studies have already begun.
For example, BBO is designed to detect inflationary generated  gravitational wave background with amplitude
$\Omega_{GW}\lsim  10^{-16}$ around 1 Hz where the white dwarf binary  confusion noise is
absent.  
The relevant
frequency for PBH search ($f \lsim 10^{-5}$Hz) is largely different from
1 Hz.  However, if the mission design is such that the proof-mass noise can be controlled well below
$\sim 10^{-5}$Hz, such a mission would become a powerful tool for PBH  search.

{\it Acknowledgments:} We thank members of the Caltech TAPIR group for
helpful comments. This work has been
supported by the Sherman Fairchild foundation, DOE DE-FG 03-92-ER40701 at Caltech
(AC) and NASA grant NAG5-10707 (NS).

\end{document}